\documentclass[a4paper,11pt,fleqn]{article}

\usepackage{amsmath,amssymb,amsthm} 

\allowdisplaybreaks

\newcommand{\Inv}{\mathop{\rm Inv}\nolimits}

\newcommand{\NilRad}{\mathop{\rm NR}\nolimits}
\newcommand{\diag}{\mathop{\rm diag}\nolimits}

\newcommand{\todo}[1][\null]{\ensuremath{\clubsuit}}

\newtheorem{theorem}{Theorem}
\newtheorem{lemma}[theorem]{Lemma}
\newtheorem{corollary}[theorem]{Corollary}

{\theoremstyle{definition}
\newtheorem{remark}[theorem]{Remark}
}

\begin{document}


\par\noindent {\LARGE\bf 
Invariants of triangular Lie algebras\\ with one nilindependent diagonal element
\par}

\vspace{4mm}\par\noindent {\large 
Vyacheslav Boyko~$^\dag$, Jiri Patera~$^\ddag$ and Roman O. Popovych~$^{\dag\S}$
} \par\vspace{2mm}\par

\vspace{2mm}\par\noindent {\it
$^\dag$~Institute of Mathematics of NAS of Ukraine, 3
Tereshchenkivs'ka Str., Kyiv, 01004 Ukraine}\\
$\phantom{^\dag}$~E-mail: boyko@imath.kiev.ua, rop@imath.kiev.ua\par

\vspace{2mm}\par\noindent {\it
$^\ddag$~Centre de Recherches Math\'ematiques,
Universit\'e de Montr\'eal,\\
$\phantom{^\ddag}$~C.P. 6128 succursale Centre-ville, Montr\'eal (Qu\'ebec), H3C 3J7 Canada}\\
$\phantom{^\ddag}$~E-mail: patera@CRM.UMontreal.CA
\par

\vspace{2mm}\par\noindent {\it
\noindent $^\S$~Wolfgang Pauli Institut, Oskar-Morgenstern-Platz 1, 1090 Wien, Austria\par
}

\vspace{5mm}\par\noindent\hspace*{10mm}\parbox{145mm}{\small
The invariants of solvable triangular Lie algebras with one nilindependent diagonal element 
are studied exhaustively. 
Bases of the invariant sets of all such algebras are constructed using an original algebraic algorithm
based on Cartan's method of moving frames and the special technique developed for triangular and related algebras
in [{\it J.\,Phys.\,A: Math.\,Theor.} {\bf 40} (2007), 7557--7572].
The conjecture of Tremblay and Winternitz [{\it J.\,Phys.\,A: Math.\,Gen.} {\bf 34} (2001), 9085--9099] on the number
and form of elements in the bases is completed and proved.
}\par\vspace{3mm}

\section{Introduction}

The possibility of finding complete explicit formulae for the
invariants of a Lie algebra is without a doubt connected with some
precise knowledge of its structure. Since the invariants of Lie
algebras are their essential characteristics, and are important in
their application, the exhaustive description of invariants was
attempted for all known structures of Lie algebras.

\looseness=-1
This problem was solved in the cases of the semi-simple and
low-dimensional Lie algebras, for physically relevant Lie algebras
of fixed dimensions, as well as Lie algebras with the simplest (Abelian) radicals (see, e.g., references
in~\cite{Boyko&Patera&Popovych2006,Campoamor-Stursberg2007,Patera&Sharp&Winternitz1976}).
Further progress in the study of Lie algebra invariants (also called
generalized Casimir operators) are closely related with progress in
the classification of classes of solvable algebras and unsolvable
Lie algebras with non-trivial radicals of arbitrary finite dimensions
\cite{Ancochea&Campoamor-Stursberg&GarciaVergnolle2006,Campoamor-Stursberg2006a,Campoamor-Stursberg2007,Ndogmo2004,
Ndogmo&Wintenitz1994a,Ndogmo&Wintenitz1994b,Rubin&Winternitz1993,Snobl&Winternitz2005,Tremblay&Winternitz1998,
Tremblay&Winternitz2001}. 
The infinitesimal method became the convention for the computation of invariants. 
It is based on the integration of a linear system of first-order partial differential
equations associated with infinitesimal operators of coadjoint action. 
Algebraic tools were occasionally applied in the construction of invariants for special classes of algebras 
\cite{Kaneta1984a,Perroud1983}.

In~\cite{Boyko&Patera&Popovych2006,Boyko&Patera&Popovych2007} an
original pure algebraic approach to invariants was proposed and developed. 
It involves Fels--Olver's approach to Cartan's method of
moving frames~\cite{Fels&Olver1998,Fels&Olver1999}. 
(For modern development of the moving frames method and more references see also
\cite{Olver&Pohjanpelto2007}). More precisely, the technique of the
moving frames method is specialized in its frameworks for the case
of coadjoint action of the associated inner automorphism groups on
the dual spaces of Lie algebras. Unlike the infinitesimal method,
such an approach allows us to avoid solving systems of differential
equations, replacing them by algebraic equations. 
As a result, it is essentially simpler to apply.

Different versions of the algebraic approach were tested
in~\cite{Boyko&Patera&Popovych2006,Boyko&Patera&Popovych2007} for
the Lie algebras of dimensions not greater than 6 and a wide range
of known solvable Lie algebras of arbitrary finite dimensions with a
fixed structure of nilradicals. A special technique for working with
solvable Lie algebras having triangular nilradicals was developed
in~\cite{Boyko&Patera&Popovych2007b}. Fundamental invariants were
constructed with this technique for the algebras $\mathfrak t_0(n)$,
$\mathfrak t(n)$ and $\mathfrak{st}(n)$. Here $\mathfrak t_0(n)$
denotes the nilpotent Lie algebra of strictly upper triangular $n\times n$ matrices over the field $\mathbb F$, 
where $\mathbb F$ is either $\mathbb C$ or $\mathbb R$. 
The solvable Lie algebras of non-strictly upper triangular and special upper triangular $n\times n$ matrices 
are denoted by $\mathfrak t(n)$ and $\mathfrak{st}(n)$, respectively.

The invariants of triangular algebras were first considered
in~\cite{Tremblay&Winternitz2001}, with the infinitesimal method.
Theorem~1 on the Casimir operators of $\mathfrak t_0(n)$ and
Proposition~1 on the invariants of $\mathfrak{st}(n)$
from~\cite{Tremblay&Winternitz2001} were completely corroborated
in~\cite{Boyko&Patera&Popovych2007b}. Note that Proposition~1 was
only a conjecture derived after the calculation of the invariants
for all partial values $n\leqslant13$. Another conjecture was
formulated in~\cite{Tremblay&Winternitz2001} as Proposition~2 on 
invariant bases of 
solvable Lie algebras having $\mathfrak t_0(n)$ as their nilradicals
and possessing a minimal (one) number of nilindependent `diagonal'
elements. It was invented after the construction of the invariants
for a narrower range of~$n$ than in the case of $\mathfrak{st}(n)$ 
(namely, $n\leqslant8$), and it has not been proved as of this writing. 
In the framework of the infinitesimal approach, the necessary calculations are too
cumbersome, even more so for these algebras. This probably led to
the reduction of possibility of computational experiments and to the impossibility
of proving the aforementioned conjectures for arbitrary values of~$n$.

In this paper we rigorously construct bases of the invariant sets
for all the solvable Lie algebras with nilradicals isomorphic to
$\mathfrak t_0(n)$ and one nilindependent `diagonal' element for
arbitrary relevant values of~$n$ (i.e., $n>1$). We use the algebraic
approach along with some additional technical tools that were
developed for triangular and related algebras
in~\cite{Boyko&Patera&Popovych2007b}. All the steps of the algorithm
are implemented one after another: construction of the coadjoint
representation of the corresponding Lie group and its fundamental
lifted invariant
(Section~\ref{SectionRepresentationOfCoadjointAction}), excluding
the group parameters from the lifted invariants by the normalization
procedure that results in a basis of the invariants for the
coadjoint action (Section~\ref{SectionInvariantsOfCoadjointAction})
and re-writing this basis as a basis of the invariants of the Lie
algebra under consideration (Section~\ref{SectionInvariants}).
Description of some necessary notions and statements, precise
formulation and discussion of the technical details of the applied
algorithm can be found
in~\cite{Boyko&Patera&Popovych2006,Boyko&Patera&Popovych2007,Boyko&Patera&Popovych2007b},
and hence are omitted here. The calculations involved in any step
are more complicated than in~\cite{Boyko&Patera&Popovych2007b}, but
due to optimization they remain quite useful. 
There are two cases, depending on the parameters of the algebra, 
that differ in the necessary number of normalization constraints and, 
therefore, in the cardinality of the fundamental invariants. 
The conjecture given in Proposition~2 of~\cite{Tremblay&Winternitz2001} is completed and proved.

\section{Representation of the coadjoint action}\label{SectionRepresentationOfCoadjointAction}

Let the underlying field~$\mathbb F$ be either $\mathbb C$ or $\mathbb R$.
Consider the solvable Lie algebra $\mathfrak t_\gamma(n)$ 
with the nilradical $\NilRad(\mathfrak t_\gamma(n))$ isomorphic to $\mathfrak t_0(n)$ 
and one nilindependent element $f$, which acts on elements of the nilradical in the same way
as the diagonal matrix $\Gamma=\diag(\gamma_1,\dots,\gamma_n)$ acts on strictly triangular matrices,
also consider $\Gamma$ as being a matrix non-proportional to the identity matrix.
The tuple $\gamma=(\gamma_1,\dots,\gamma_n)$ has different elements due to the condition on $\Gamma$.
It is defined up to a nonzero multiplier, a homogeneous shift of entry values
and the mirror reflection with respect to the central vertical line.
In other words, the algebras $\mathfrak t_\gamma(n)$ and $\mathfrak t_{\gamma'}(n)$ are isomorphic if and only if
there exist $\lambda,\mu\in\mathbb F$ with $\lambda\ne0$ such that 
\[
\gamma'_i=\lambda\gamma_i+\mu,\quad i=1,\dots,n, \qquad\mbox{or}\qquad \gamma'_i=\lambda\gamma_{n-i+1}+\mu,\quad i=1,\dots,n.
\]
The tuples $\gamma$ and $\gamma'$ are assumed to be equivalent.
Up to this equivalence, the additional condition $\mathop{\rm tr}\Gamma=\sum_i\gamma_i=0$ can be 
imposed on the algebra parameters. Therefore, the algebra $\mathfrak t_\gamma(n)$ is naturally 
embedded into $\mathfrak{st}(n)$ as an ideal, thus identifying $\NilRad(\mathfrak t_\gamma(n))$ with 
$\mathfrak t_0(n)$ and $f$ with $\Gamma$.

The concatenation of the canonical basis of $\NilRad(\mathfrak t_\gamma(n))$ and the singleton~$(f)$
is chosen as the canonical basis of $\mathfrak t_\gamma(n)$.
In the basis of $\NilRad(\mathfrak t_\gamma(n))$ we use a `matrix' enumeration of the basis elements
$e_{ij}$, $i<j$, with an `increasing' pair of indices, similarly to
the canonical basis $(E^n_{ij},\,i<j)$ of the isomorphic matrix algebra $\mathfrak t_0(n)$.

Hereafter
$E^n_{ij}$ (for fixed values $i$ and $j$) denotes the $n\times n$ matrix $(\delta_{ii'}\delta_{jj'})$
with $i'$ and $j'$ running the numbers of rows and columns, respectively,
i.e., the $n\times n$ matrix with the unit on the cross of the $i$th row and the $j$th column and zero otherwise.
The indices $i$, $j$, $k$ and $l$ run at most from 1 to~$n$.
Only additional constraints on the indices are indicated.

Thus, the basis elements $e_{ij}\sim E^n_{ij}$, $i<j$, and $f\sim\sum_i \gamma_iE^n_{ii}$ satisfy the commutation relations
\[
[e_{ij},e_{i'\!j'}]=\delta_{i'\!j}e_{ij'}-\delta_{ij'}e_{i'\!j}, \quad
[f,e_{ij}]=(\gamma_i-\gamma_j)e_{ij}, \quad
\]
where $\delta_{ij}$ is the Kronecker delta.

The Lie algebra $\mathfrak t_\gamma(n)$ can be considered as the Lie algebra of the Lie subgroup
\[
{\rm T}_\gamma(n)=\{B\in {\rm T}(n)\mid \exists \, \varepsilon\in\mathbb F\colon b_{ii}=e^{\gamma_i\varepsilon}\}
\]
of the Lie group ${\rm T}(n)$ of non-singular upper triangular $n\times n$ matrices.

Let $e_{ji}^*$, $x_{ji}$ and $y_{ij}$ denote
the basis element and the coordinate function in the dual space $\mathfrak t_\gamma^*(n)$ and
the coordinate function in~$\mathfrak t_\gamma(n)$,
which correspond to the basis element~$e_{ij}$, $i<j$.
In~particular, $\langle e_{j'\!i'}^*,e_{ij}\rangle=\delta_{ii'}\delta_{jj'}.$
The reverse order of subscripts of the dual elements and coordinates
is justified by the simplification of a matrix representation of lifted invariants.
$f^*$, $x_0$ and $y_0$ denote the similar objects corresponding to the basis element~$f$.
We additionally set $y_{ii}=\gamma_iy_0$ and then
complete the collections of~$x_{ji}$ and of~$y_{ij}$ with zeros to the matrices $X$ and $Y$.
Hence $X$ is a strictly lower triangular matrix and $Y$ is a non-strictly upper triangular one.
The analogous `matrix' with the significant elements~$e_{ij}$, $i<j$, is denoted by $\mathcal E$.

\begin{lemma}\label{LemmaOnLiftedInvsOfSolvAlgsWithTriangularNilradicalAnd1NonnilpElement}
A complete set of functionally independent lifted invariants of ${\rm Ad}^*_{{\rm T}_\gamma(n)}$
is exhausted by the expressions
\[
\mathcal I_{ij}=\sum_{i\leqslant i',\,j'\leqslant j}b_{ii'}\widehat b_{j'\!j}x_{i'\!j'}, \quad j<i,
\qquad
\mathcal I_0=x_0+\sum_{j<i}\,\sum_{j\leqslant l\leqslant i}\gamma_lb_{li}\widehat b_{jl}x_{ij},
\]
where
$B=(b_{ij})$ is an arbitrary matrix from ${\rm T}_\gamma(n)$, and
$B^{-1}=(\widehat b_{ij})$ is the inverse matrix of~$B$.
\end{lemma}

\begin{proof} The adjoint action of $B\in{\rm T}_\gamma(n)$ on the matrix~$Y$ is
${\rm Ad}_BY=BYB^{-1}$, i.e.,
\[
{\rm Ad}_B\biggl(y_0f+\sum_{i<j}y_{ij}e_{ij}\biggr)
=y_0f+y_0\sum_{i<j}\,\sum_{i\leqslant i'\leqslant j}b_{ii'}\gamma_{i'}\widehat b_{i'\!j}e_{ij}
+\sum_{i\leqslant i'<j'\leqslant j}b_{ii'}y_{i'\!j'}\widehat b_{j'\!j}e_{ij}.
\]
After changing $e_{ij}\to x_{ji}$, $y_{ij}\to e_{ji}^*$, $f\to x_0$, $y_0\to f^*$,
$b_{ij}\leftrightarrow \widehat b_{ij}$
in the latter equality, we obtain the representation for the coadjoint action of~$B$
\begin{gather*}
{\rm Ad}_B^*\biggl(x_0f^*+\sum_{i<j}x_{ji}e_{ji}^*\biggr)
=x_0f^*+\sum_{i<j}\,\sum_{i\leqslant i'\leqslant j}b_{i'\!j}x_{ji}\widehat b_{ii'}\gamma_{i'\!}f^*
+\sum_{i\leqslant i'<j'\leqslant j}b_{j'\!j}x_{ji}\widehat b_{ii'}e_{j'\!i'}^*\\
\qquad
=\biggl(x_0+\sum_{i<j}\,\sum_{i\leqslant i'\leqslant j}b_{i'\!j}x_{ji}\widehat b_{ii'}\gamma_{i'\!}\biggr)f^*
+\sum_{i'<j'}(BXB^{-1})_{j'\!i'}e_{j'\!i'}^*.
\end{gather*}
Therefore, $\mathcal I_0$ and the elements $\mathcal I_{ij}$, $j<i$, of the matrix $\mathcal I=BXB^{-1}$, where $B\in{\rm T}_\gamma(n)$,
form a fundamental lifted invariant of ${\rm Ad}^*_{{\rm T}_\gamma(n)}$.
\end{proof}

\begin{remark}
The complete set of parameters in the above representation of lifted invariants is formed by $b_{ij}$, $j<i$, and $\varepsilon$.
The center of the group ${\rm T}_\gamma(n)$ is nontrivial only if $\gamma_1=\gamma_n$, namely,
then $\mathrm Z({\rm T}_\gamma(n))=\{E^n+b_{1n}E^n_{1n},\ b_{1n}\in\mathbb F\}$. 
Here $E^n=\diag(1,\ldots,1)$ is the $n\times n$ identity matrix.
In this case the inner automorphism group of~$\mathfrak t_\gamma(n)$ is isomorphic to the factor-group
${\rm T}_\gamma(n)/\mathrm Z({\rm T}_\gamma(n))$ and hence its dimension is $\frac12n(n-1)$.
The parameter $b_{1n}$ in the representation of the lifted invariants is thus inessential.
Otherwise, the inner automorphism group of~$\mathfrak t_\gamma(n)$ is isomorphic to the  whole group ${\rm T}_\gamma(n)$,
and all the parameters in the constructed lifted invariants are essential.
\end{remark}

\section{Invariants of the coadjoint action}\label{SectionInvariantsOfCoadjointAction}

Below $A^{i_1,i_2}_{j_1,j_2}$, where $i_1\leqslant i_2$, $j_1\leqslant j_2$,
denotes the submatrix $(a_{ij})^{i=i_1,\ldots,i_2}_{j=j_1,\ldots,j_2}$ of a matrix $A=(a_{ij})$.
The standard notation $|A|=\det A$ is used.
The conjugate value of $k$ with respect to $n$ is denoted by $\varkappa$, i.e., $\varkappa=n-k+1$.

At first we formulate the technical lemma from \cite{Boyko&Patera&Popovych2007b}, applied to the proof of the following theorem.

\begin{lemma}\label{LemmaOnEqualitiesWithSubmatrix}
Suppose $1<k<n$.
If $|X^{\varkappa+1,n}_{1,k-1}|\ne0$, then for any $\beta\in\mathbb F$
\begin{gather*}\arraycolsep=0.5ex
\beta-X^{i,i}_{1,k-1}(X^{\varkappa+1,n}_{1,k-1})^{-1}X^{\varkappa+1,n}_{j,j}=
\frac{(-1)^{k+1}}{|X^{\varkappa+1,n}_{1,k-1}|}
\left|\begin{array}{lc} X^{i,i}_{1,k-1} & \beta \\[1ex]
X^{\varkappa+1,n}_{1,k-1}& X^{\varkappa+1,n}_{j,j} \end{array}\!\right|.
\end{gather*}
In particular,
$x_{\varkappa k}-X^{\varkappa,\varkappa}_{1,k-1}(X^{\varkappa+1,n}_{1,k-1})^{-1}X^{\varkappa+1,n}_{k,k}=
(-1)^{k+1} |X^{\varkappa+1,n}_{1,k-1}|^{-1} |X^{\varkappa,n}_{1,k}|$.
Analogously
\begin{gather*}
\left(x_{\varkappa j}-X^{\varkappa,\varkappa}_{1,k-1}(X^{\varkappa+1,n}_{1,k-1})^{-1}X^{\varkappa+1,n}_{j,j}\right)
\left(x_{jk}-X^{j,j}_{1,k-1}(X^{\varkappa+1,n}_{1,k-1})^{-1}X^{\varkappa+1,n}_{k,k}\right)
\\[1ex]\arraycolsep=.5ex
\qquad=\frac{1}{|X^{\varkappa+1,n}_{1,k-1}|}
\left|\begin{array}{lc} X^{j,j}_{1,k} & \beta \\[1ex] X^{\varkappa,n}_{1,k}& X^{\varkappa,n}_{j,j} \end{array}\!\right|+
\frac{|X^{\varkappa,n}_{1,k}|}{|X^{\varkappa+1,n}_{1,k-1}|^2}
\left|\begin{array}{lc} X^{j,j}_{1,k-1} & \beta \\[1ex] X^{\varkappa+1,n}_{1,k-1}& X^{\varkappa+1,n}_{j,j} \end{array}\!\right|.
\end{gather*}
\end{lemma}

\begin{theorem}\label{TheoremOnBasisOfInvsOfCoadjRepresentationOfSolvAlgsWithTriangularNilradicalAnd1NonnilpElement}
A basis of $\Inv({\rm Ad}^*_{{\rm T}_\gamma(n)})$ consists of the expressions
\[
1)\ |X^{\varkappa,n}_{1,k}|, \quad k=1, \ldots, \left[\frac n2\right], \qquad
x_0+\sum_{k=1}^{\left[\frac n2\right]} \frac{(-1)^{k+1}}{|X^{\varkappa,n}_{1,k}|} (\gamma_k-\gamma_{k+1}) \sum_{k<i<\varkappa}
\left|\begin{array}{lc} X^{i,i}_{1,k} & 0 \\[1ex] X^{\varkappa,n}_{1,k}& X^{\varkappa,n}_{i,i} \end{array}\!\right|
\]
if $\gamma_k=\gamma_\varkappa$ for all $k\in\{1,\dots,[n/2]\}$ or of the expressions
\begin{gather*}
2)\ |X^{\varkappa,n}_{1,k}|, \quad k=1, \ldots, k_0-1, \qquad
|X^{\varkappa_0,n}_{1,k_0}|^{\alpha_k}|X^{\varkappa,n}_{1,k}|,
\quad k=k_0+1,\ldots,\left[\frac n2\right]
\end{gather*}
otherwise. Here $k_0$ is the minimal value of $k$ for which $\gamma_k\ne\gamma_\varkappa$ and
\[
\alpha_k=-\sum_{i=k_0}^k\frac{\gamma_{n-i+1}-\gamma_i}{\gamma_{n-k_0+1}-\gamma_{k_0}}.
\]
\end{theorem}

\begin{remark}\label{RemarkOnDomainOfInvsAnsPowers}
In general, expressions in Theorem~\ref{TheoremOnBasisOfInvsOfCoadjRepresentationOfSolvAlgsWithTriangularNilradicalAnd1NonnilpElement} 
are not defined on the whole space~$\mathfrak t_\gamma^*(n)$. 
Some singularities can be removed by recombining these expressions. 
In particular, the last expression of the first case is well defined 
only if $|X^{\varkappa,n}_{1,k}|\ne0$ for all $k\in\{1,\dots,[n/2]\}$ with $\gamma_k-\gamma_{k+1}\ne0$. 
Multiplying it by the product of $|X^{\varkappa,n}_{1,k}|$ with $k\in\{1,\dots,[n/2]\}$ and $\gamma_k-\gamma_{k+1}\ne0$, 
we obtain a polynomial in~$x$'s, which is defined on the whole space~$\mathfrak t_\gamma^*(n)$. 
The second case is more complicated. 
If $k_0<[n/2]$ and some of the exponents~$\alpha_k$'s are not integer, then for $\mathbb F=\mathbb C$
a branch of the ln should be fixed and then used for expressing, 
via the exponential function, all powers involved in the expressions of the second case. 
If the underlying field is real, these powers are defined for any values of their exponents only for~$x$'s, 
where the determinants being their bases are positive. 
In the general situation with the real field, when an exponent is not an integer or a rational number with odd denominator, 
the corresponding determinant should be replaced by its absolute value. 
A polynomial (and hence, globally defined) basis of invariants exists in the second case only if 
either $n\in2\mathbb N$ and $k_0=n/2$ 
or $\alpha_k=0$ for all $k\in K:=\{k_0+1,\dots,[n/2]\}$ 
or $\alpha_k\in\mathbb Q$ for all $k\in K$ and $\alpha_k>0$ for some $k\in K$.
\end{remark}

\begin{proof}
Under normalization we impose the following restriction on the lifted invariants $\mathcal I_{ij}$, $j<i$:
\[
\mathcal I_{ij}=0 \quad\mbox{if}\quad j<i,\ (i,j)\not=(n-j'+1,j'),\ j'=1,\ldots,\left[\frac{n}2\right].
\]
This means we do not only fix the values of the elements of the lifted invariant matrix~$\mathcal I$,
which are situated on the secondary diagonal under the main diagonal.
The other significant elements of~$\mathcal I$ are put equal to 0.

The decision on what to do with the singular lifted invariant~$\mathcal I_0$ and the
secondary-diagonal lifted invariants $\mathcal I_{\varkappa k}$,  $k=1,\dots,[n/2]$, is left for later, 
since it turns out that the necessity of imposing normalization conditions on them depends on the values of~$\gamma$. 
As shown below, the final normalization in all the cases provides satisfying the conditions of 
Proposition~1 from~\cite{Boyko&Patera&Popovych2007b} and, therefore, is correct.

\looseness=-1
In view of the (triangular) structure of the matrices $B$ and $X$,
the formula $\mathcal I=BXB^{-1}$ determining the matrix part of the lifted invariants implies $BX=\mathcal IB$. 
This matrix equality is also significant for the matrix elements underlying the main diagonals
of the left and right hand sides, i.e.,
\[
e^{\gamma_i\varepsilon}x_{ij}+\sum_{i<i'}b_{ii'}x_{i'\!j}=\mathcal I_{ij}e^{\gamma_j\varepsilon}+\sum_{j'<j}\mathcal I_{ij'}b_{j'\!j}, \quad j<i.
\]
For convenience the latter system is divided under the chosen normalization conditions into four sets of subsystems
\begin{gather*}
S_1^k\colon\qquad e^{\gamma_\varkappa\varepsilon}x_{\varkappa j}+\sum_{i'>\varkappa}b_{\varkappa i'}x_{i'\!j}=0, \qquad
i=\varkappa,\quad j<k,\quad k=2,\ldots,\left[\frac{n+1}2\right],
\\
S_2^k\colon\qquad
e^{\gamma_\varkappa\varepsilon}x_{\varkappa k}+\sum_{i'>\varkappa}b_{\varkappa i'}x_{i'\!k}=\mathcal I_{\varkappa k}e^{\gamma_k\varepsilon}, \qquad
i=\varkappa,\quad j=k,\quad k=1,\ldots,\left[\frac n2\right],
\\
S_3^k\colon\qquad e^{\gamma_\varkappa\varepsilon}x_{\varkappa j}+\sum_{i'>\varkappa}b_{\varkappa i'}x_{i'\!j}=\mathcal I_{\varkappa k}b_{kj}, \qquad
i=\varkappa,\quad k<j<\varkappa,\quad k=1,\ldots,\left[\frac n2\right]-1,
\\
S_4^k\colon\qquad e^{\gamma_k\varepsilon}x_{kj}+\sum_{i'>k}b_{ki'}x_{i'\!j}=0, \qquad
i=k,\quad j<k,\quad k=2,\ldots,\left[\frac n2\right],
\end{gather*}
and solve them one after another.
The subsystem~$S_2^1$ consists of the single equation
\[
\mathcal I_{n1}=x_{n1}e^{(\gamma_n-\gamma_1)\varepsilon}.
\]
For any fixed $k\in\{2,\dots,[n/2]\}$ the subsystem $S_1^k \cup S_2^k$ is a well-defined system of linear equations with respect to
$b_{\varkappa i'}$, $i'>\varkappa$, and $\mathcal I_{\varkappa k}$.
Analogously, the subsystem  $S_1^k$ for $k=\varkappa=[(n+1)/2]$ in the case of an odd~$n$
is a well-defined system of linear equations with respect to $b_{k i'}$, $i'>k$.
The solutions of the above subsystems are expressions of $x_{i'\!j}$, $i'\geqslant\varkappa$, $j<k$, and $\varepsilon$:
\begin{gather*}
\mathcal I_{\varkappa k}=(-1)^{k+1}
\frac{|X^{\varkappa,n}_{1,k}|}{|X^{\varkappa+1,n}_{1,k-1}|}\,e^{(\gamma_\varkappa-\gamma_k)\varepsilon},
\quad k=2, \ldots, \left[\frac n2\right],
\\
B^{\varkappa,\varkappa}_{\varkappa+1,n}=-e^{\gamma_\varkappa\varepsilon}X^{\varkappa,\varkappa}_{1,k-1}(X^{\varkappa+1,n}_{1,k-1})^{-1},
\quad k=2, \ldots, \left[\frac {n+1}2\right].
\end{gather*}

After substituting the expressions of $\mathcal I_{\varkappa k}$ and $b_{\varkappa i'}$, $i'>\varkappa$, via $\varepsilon$ and $x$'s
into $S_3^k$, we trivially resolve $S_3^k$ with respect to $b_{kj}$ as an uncoupled system of linear equations:
\begin{gather*}
b_{1j}=e^{\gamma_1\varepsilon}\frac{x_{nj}}{x_{n1}}, \quad
1<j<n,
\\
b_{kj}=(-1)^{k+1}e^{\gamma_k\varepsilon}
\frac{|X^{\varkappa+1,n}_{1,k-1}|}{|X^{\varkappa,n}_{1,k}|}
\left(x_{\varkappa j}-X^{\varkappa,\varkappa}_{1,k-1}(X^{\varkappa+1,n}_{1,k-1})^{-1}X^{\varkappa+1,n}_{j,j}\right)
=\frac{e^{\gamma_k\varepsilon}}{|X^{\varkappa,n}_{1,k}|}
\arraycolsep=0.5ex
\left|\begin{array}{ll} X^{\varkappa,\varkappa}_{1,k-1} & x_{\varkappa j} \\[1ex]
X^{\varkappa+1,n}_{1,k-1}& X^{\varkappa+1,n}_{j,j} \end{array}\!\right|,
\\
k<j<\varkappa,\quad k=2,\ldots,\left[\frac n2\right]-1.
\end{gather*}

Performing the subsequent substitution of the calculated expressions for $b_{kj}$ into $S_4^k$, for any fixed appropriate~$k$
we obtain a well-defined system of linear equations, e.g., with respect to $b_{ki'}$, $i'>\varkappa$.
Its solution is expressed via $x$'s, $b_{k\varkappa}$ and $\varepsilon$:
\begin{gather*}
B^{k,k}_{\varkappa+1,n}=-\biggl(e^{\gamma_k\varepsilon}X^{k,k}_{1,k-1}+
\sum_{k< j\leqslant\varkappa}b_{kj}X^{j,j}_{1,k-1}\biggr)(X^{\varkappa+1,n}_{1,k-1})^{-1}
\\\phantom{B^{k,k}_{\varkappa+1,n}}
=-b_{k\varkappa}X^{\varkappa,\varkappa}_{1,k-1}(X^{\varkappa+1,n}_{1,k-1})^{-1}
-\frac{e^{\gamma_k\varepsilon}}{|X^{\varkappa,n}_{1,k}|}
\sum_{k\leqslant j<\varkappa}\arraycolsep=0.5ex
\left|\begin{array}{ll} X^{\varkappa,\varkappa}_{1,k-1} & x_{\varkappa j} \\[1ex]
X^{\varkappa+1,n}_{1,k-1}& X^{\varkappa+1,n}_{j,j} \end{array}\!\right|
X^{j,j}_{1,k-1}(X^{\varkappa+1,n}_{1,k-1})^{-1},
\\
k=2,\ldots,\left[\frac n2\right].
\end{gather*}

The expression of the lifted invariant~$\mathcal I_0$ is rewritten, taking into account the already imposed normalization constraints
(note that $\varkappa=[(n+1)/2]+1$ if $k=[n/2]$):
\begin{gather*}
\mathcal I_0=x_0
+\sum_l\gamma_l\widehat b_{ll}\sum_{l<i}b_{li}x_{il}
+\sum_{k=2}^{\left[\frac{n+1}2\right]}\sum_{j<k}\gamma_k\widehat b_{jk}\sum_{i\geqslant k}b_{ki}x_{ij}
+\sum_{k=1}^{\left[\frac n2\right]}\Biggl(\,\sum_{j<k}+\sum_{k\leqslant j<\varkappa}\,\Biggr)
\gamma_\varkappa\widehat b_{j\varkappa}\sum_{i\geqslant \varkappa}b_{\varkappa i}x_{ij}
\\\phantom{\mathcal I_0}
=x_0
+\sum_l\gamma_l\widehat b_{ll}\sum_{l<i}b_{li}x_{il}
+\sum_{k=1}^{\left[\frac n2\right]}\gamma_\varkappa\mathcal I_{\varkappa k}\sum_{k\leqslant j<\varkappa}b_{kj}\widehat b_{j\varkappa}
\\\phantom{\mathcal I_0}
=x_0
+\sum_{k=1}^{\left[\frac n2\right]}\gamma_k\widehat b_{kk}\Biggl(\,\sum_{k<i\leqslant\varkappa}+\sum_{i>\varkappa}\,\Biggr)b_{ki}x_{ik}
+\sum_{k=1}^{\left[\frac{n+1}2\right]}\gamma_\varkappa\widehat b_{\varkappa\varkappa}\sum_{i>\varkappa}b_{\varkappa i}x_{i\varkappa}
-\sum_{k=1}^{\left[\frac n2\right]}\gamma_\varkappa\widehat b_{\varkappa\varkappa}\mathcal I_{\varkappa k}b_{k\varkappa}.
\end{gather*}
Then the found expressions for $b$'s and $I_{\varkappa k}$ are substituted into the derived expression of~$\mathcal I_0$:
\begin{gather*}
\mathcal I_0=x_0
+\gamma_1e^{-\gamma_1\varepsilon}\sum_{1<i\leqslant n}b_{1i}x_{i1}
+\sum_{k=2}^{\left[\frac n2\right]}\gamma_ke^{-\gamma_k\varepsilon} \sum_{k<i\leqslant\varkappa}b_{ki}
\left(x_{ik}-X^{i,i}_{1,k-1}(X^{\varkappa+1,n}_{1,k-1})^{-1}X^{\varkappa+1,n}_{k,k}\right)
\\\phantom{\mathcal I_0=}
-\sum_{k=2}^{\left[\frac n2\right]}\gamma_kX^{k,k}_{1,k-1}(X^{\varkappa+1,n}_{1,k-1})^{-1}X^{\varkappa+1,n}_{k,k}
+\sum_{k=1}^{\left[\frac{n+1}2\right]}\gamma_\varkappa\widehat b_{\varkappa\varkappa}\sum_{i>\varkappa}b_{\varkappa i}x_{i\varkappa}
-\sum_{k=1}^{\left[\frac n2\right]}\gamma_\varkappa\widehat b_{\varkappa\varkappa}\mathcal I_{\varkappa k}b_{k\varkappa}
\\\phantom{\mathcal I_0}
=x_0
+(\gamma_1-\gamma_n)e^{-\gamma_1\varepsilon}b_{1n}x_{n1}
+\sum_{k=2}^{\left[\frac n2\right]}(\gamma_k-\gamma_\varkappa)e^{-\gamma_k\varepsilon}b_{k\varkappa}(-1)^{k+1}
\frac{|X^{\varkappa,n}_{1,k}|}{|X^{\varkappa+1,n}_{1,k-1}|}
\\\phantom{\mathcal I_0=}
-\sum_{k=2}^{\left[\frac n2\right]}
\gamma_kX^{k,k}_{1,k-1}(X^{\varkappa+1,n}_{1,k-1})^{-1}X^{\varkappa+1,n}_{k,k}
-\sum_{k=2}^{\left[\frac{n+1}2\right]}
\gamma_\varkappa X^{\varkappa,\varkappa}_{1,k-1}(X^{\varkappa+1,n}_{1,k-1})^{-1}X^{\varkappa+1,n}_{\varkappa,\varkappa}
\\\phantom{\mathcal I_0=}\arraycolsep=.5ex
+\sum_{k=1}^{\left[\frac n2\right]} \frac{(-1)^{k+1}\gamma_k}{|X^{\varkappa,n}_{1,k}|}\sum_{k<i<\varkappa}
\left|\begin{array}{lc} X^{i,i}_{1,k} & 0 \\[1ex] X^{\varkappa,n}_{1,k}& X^{\varkappa,n}_{i,i} \end{array}\!\right|
+\sum_{k=2}^{\left[\frac n2\right]} \frac{(-1)^{k+1}\gamma_k}{|X^{\varkappa+1,n}_{1,k-1}|}\sum_{k<i<\varkappa}
\left|\begin{array}{lc} X^{i,i}_{1,k-1} & 0 \\[1ex] X^{\varkappa+1,n}_{1,k-1}& X^{\varkappa+1,n}_{i,i} \end{array}\!\right|.
\end{gather*}

If $\gamma_k=\gamma_\varkappa$ for all $k\in\{1,\dots,[n/2]\}$, then
$\mathcal I_{\varkappa k}$, $k=1,\dots,[n/2]$, and $\mathcal I_0$ do not depend on the parameters $b$ and $\varepsilon$, i.e.,
they are invariants. For a basis to be simpler, $\hat{\mathcal I}_0=\mathcal I_0$ is taken, as well as $\hat{\mathcal I}_1=\mathcal I_{n1}$
and the combinations $\hat{\mathcal I}_k=(-1)^{k+1}\mathcal I_{\varkappa k}\hat{\mathcal I}_{k-1}$, $k=2,\dots,[n/2]$,
resulting in the first tuple of invariants from the statement of the theorem.
Let us show that the above formula for $\mathcal I_0$ gives exactly the expression from the statement of the theorem.
Under the supposition on $\gamma$ and after permuting terms, this formula is transformed into
\begin{gather*}\arraycolsep=.5ex
\mathcal I_{p0}=x_{p0}
+\sum_{k=1}^{\left[\frac n2\right]} \frac{(-1)^{k+1}\gamma_k}{|X^{\varkappa,n}_{1,k}|}\sum_{k<i<\varkappa}
\left|\begin{array}{lc} X^{i,i}_{1,k} & 0 \\[1ex] X^{\varkappa,n}_{1,k}& X^{\varkappa,n}_{i,i} \end{array}\!\right|
+\sum_{k=2}^{\left[\frac n2\right]} \frac{(-1)^{k+1}\gamma_k}{|X^{\varkappa+1,n}_{1,k-1}|}\sum_{k<i<\varkappa}
\left|\begin{array}{lc} X^{i,i}_{1,k-1} & 0 \\[1ex] X^{\varkappa+1,n}_{1,k-1}& X^{\varkappa+1,n}_{i,i} \end{array}\!\right|
\\\phantom{\mathcal I_{p0}=}
-\sum_{k=2}^{\left[\frac n2\right]}
\gamma_kX^{k,k}_{1,k-1}(X^{\varkappa+1,n}_{1,k-1})^{-1}X^{\varkappa+1,n}_{k,k}
-\left(\sum_{k=2}^{\left[\frac n2\right]}+\sum_{k=\left[\frac n2\right]+1}^{\left[\frac{n+1}2\right]}\right)
\gamma_k X^{\varkappa,\varkappa}_{1,k-1}(X^{\varkappa+1,n}_{1,k-1})^{-1}X^{\varkappa+1,n}_{\varkappa,\varkappa}.
\end{gather*}
For convenience, denote the summation complexes in the obtained formula by $\Sigma_1$, \dots, $\Sigma_5$ 
(two and three complexes in the first and second formula's rows, respectively). 
The complex~$\Sigma_5$ contains no summands (resp.\ one summand) if $n$ is even (resp.\ odd).
Applying the first part of Lemma~\ref{LemmaOnEqualitiesWithSubmatrix} for $\beta=0$, we reduce summands of $\Sigma_3$, $\Sigma_4$ and $\Sigma_5$ 
to the form similar to that of summands of $\Sigma_2$. 
We attach the modified summands to $\Sigma_2$ and thus extend the summation intervals to $k,\dots,\varkappa$ for~$i$ (using summands of~$\Sigma_3$ and~$\Sigma_4$)
and to $2,\dots,[n/2]+1$ for~$k$ (using the summand of~$\Sigma_5$ if $n$ is odd; 
the extension is not needed if $n$ is even), 
\begin{gather*}\arraycolsep=.5ex
\mathcal I_{p0}=x_{p0}
+\sum_{k=1}^{\left[\frac n2\right]} \frac{(-1)^{k+1}\gamma_k}{|X^{\varkappa,n}_{1,k}|}\sum_{k<i<\varkappa}
\left|\begin{array}{lc} X^{i,i}_{1,k} & 0 \\[1ex] X^{\varkappa,n}_{1,k}& X^{\varkappa,n}_{i,i} \end{array}\!\right|
+\sum_{k=2}^{\left[\frac n2\right]+1} \frac{(-1)^{k+1}\gamma_k}{|X^{\varkappa+1,n}_{1,k-1}|}
\sum_{k\leqslant i\leqslant\varkappa}
\left|\begin{array}{lc} X^{i,i}_{1,k-1} & 0 \\[1ex] X^{\varkappa+1,n}_{1,k-1}& X^{\varkappa+1,n}_{i,i} \end{array}\!\right|.
\end{gather*}
The shifting of the index~$k$ by $-1$ in the last sum, $k'=k-1$ and thus $\varkappa'=\varkappa+1$, changes 
the summation intervals to $1,\dots,[n/2]$ for~$k'$ and to 
$k'+1,\dots,\varkappa'-1$ for~$i$. 
The recombination of terms leads to the required expression.

Otherwise, if there exists $k_0\in\{1,\dots,[n/2]\}$ such that $\gamma_{k_0}\ne\gamma_{\varkappa_0}$, 
then $\mathcal I_0$ necessarily depends on the parameter $b_{k_0\varkappa_0}$,
which is in the expressions of $\mathcal I_{\varkappa k}$, $k=1,\dots,[n/2]$, under the already established normalization conditions.
Hence an additional normalization condition constraining $\mathcal I_0$ should be used, e.g., $\mathcal I_0=0$.
It yields an expression for $b_{k_0\varkappa_0}$ via $x$'s, other $b_{k\varkappa}$'s and $\varepsilon$.
The exact form of the latter expression is inessential.
Suppose that  $k_0$ is the minimal $k$ for which $\gamma_k\ne\gamma_\varkappa$.
$\hat{\mathcal I}_1=\mathcal I_{n1}$
and the combinations $\hat{\mathcal I}_k=(-1)^{k+1}\mathcal I_{\varkappa k}\hat{\mathcal I}_{k-1}$, $k=2,\dots,[n/2]$, are taken.
Since $\hat{\mathcal I}_{k_0}$ explicitly depends on $\varepsilon$, 
we impose one more normalization condition 
\smash{$\hat{\mathcal I}_{k_0}=1$} or \smash{$\hat{\mathcal I}_{k_0}=\mathop{\rm sgn}|X^{\varkappa_0,n}_{1,k_0}|$} 
in the complex or real case (cf.\ Remark~\ref{RemarkOnDomainOfInvsAnsPowers}), respectively,
and, using it, exclude the parameter~$\varepsilon$ from the other $\hat{\mathcal I}$'s.
As a result, we construct the second tuple of invariants from the statement of the theorem.

Under the normalization we express the non-normalized lifted invariants via $x$'s and
compute a part of the parameters $b$'s of the coadjoint action via $x$'s and the other $b$'s.
The expressions in the obtained tuples of invariants are functionally independent. 
No equations involving only $x$'s are obtained. 
In view of Proposition~1 of~\cite{Boyko&Patera&Popovych2007b}, 
this implies that the choice of normalization constraints, which depends on values of $\gamma$, is correct.
That is why the number of the found functionally independent invariants is maximal, i.e.,
they form bases of $\Inv({\rm Ad}^*_{{\rm T}_\gamma(n)})$.
\end{proof}

\begin{corollary}\label{CorollaryOnRelatedInvsOfSolvAlgsWithTriangularNilradicalAnd1NonnilpElement}
$|X^{\varkappa,n}_{1,k}|$, $k=1,\dots,[n/2]$, are functionally independent (global) relative invariants of ${\rm Ad}^*_{{\rm T}_\gamma(n)}$
for any admissible value of~$\gamma$.
\end{corollary}

Let us recall~\cite[Definition~3.30]{Olver1995} that, given a group~$G$ acting on a set~$M$, 
a function $F\colon M\to\mathbb F$ is called a (global) {\it relative invariant} of the representation of~$G$ 
if $F(g\cdot x)=\mu(g,x)F(x)$ for all $g\in G$ and $x\in M$ and some multiplier $\mu\colon G\times M\to\mathbb F$ of this representation.

\section{Algebra invariants}\label{SectionInvariants}

\begin{theorem}\label{TheoremOnBasisOfInvsOfSolvAlgsWithTriangularNilradicalAnd1NonnilpElement}
A basis of~$\Inv(\mathfrak t_\gamma(n))$ is consists of the expressions
\[\arraycolsep=.5ex
1)\ |\mathcal E^{1,k}_{\varkappa,n}|, \quad k=1, \ldots, \left[\frac n2\right], \qquad
f+\sum_{k=1}^{\left[\frac n2\right]} \frac{(-1)^{k+1}}{|\mathcal E^{1,k}_{\varkappa,n}|} (\gamma_k-\gamma_{k+1}) \sum_{i=k+1}^{n-k}
\left|\begin{array}{lc} \mathcal E^{1,k}_{i,i} & \mathcal E^{1,k}_{\varkappa,n} \\[1ex] 0 & \mathcal E^{i,i}_{\varkappa,n} \end{array}\!\right|
\]
if $\gamma_k=\gamma_\varkappa$ for all $k\in\{1,\dots,[n/2]\}$ or of the expressions
\begin{gather*}
2)\ |\mathcal E^{1,k}_{\varkappa,n}|, \quad k=1, \ldots, k_0-1, \qquad
|\mathcal E^{1,k_0}_{\varkappa_0,n}|^{\alpha_k}|\mathcal E^{1,k}_{\varkappa,n}|,
\quad k=k_0+1,\ldots,\left[\frac n2\right]
\end{gather*}
otherwise. 
Here $\varkappa:=n-k+1$;
$\mathcal E^{i_1,i_2}_{j_1,j_2}$, $i_1\leqslant i_2$, $j_1\leqslant j_2$, denotes the matrix $(e_{ij})^{i=i_1,\ldots,i_2}_{j=j_1,\ldots,j_2}$;
$k_0$ is the minimal value of $k$ for which $\gamma_k\ne\gamma_\varkappa$ and
\[
\alpha_k=-\sum_{i=k_0}^k\frac{\gamma_{n-i+1}-\gamma_i}{\gamma_{n-k_0+1}-\gamma_{k_0}}.
\]
\end{theorem}

\begin{proof}
Consider at first the invariants from
Theorem~\ref{TheoremOnBasisOfInvsOfCoadjRepresentationOfSolvAlgsWithTriangularNilradicalAnd1NonnilpElement}, 
which do not contain the variable $x_0$ corresponding to the nilindependent element $f$.
Expanding the determinants in these invariants,
we obtain expressions of $x$'s containing only such coordinate functions that
the associated basis elements commute each to other. 
Therefore, the symmetrization procedure is trivial for them.
Since $x_{ij}\sim e_{ji}$, $j<i$, hereafter it is necessary to transpose the matrices
in the obtained expressions of invariants for representation improvement. 
Finally we construct the first part of the basis of~$\Inv(\mathfrak t_\gamma(n))$ in case~1 of the statement  
and the complete basis of~$\Inv(\mathfrak t_\gamma(n))$ in case~2. 

The symmetrization procedure for the invariant with $x_0$ presented in
Theorem~\ref{TheoremOnBasisOfInvsOfCoadjRepresentationOfSolvAlgsWithTriangularNilradicalAnd1NonnilpElement} 
also can be assumed trivial. To show this, we again expand all the determinants.  
Only the monomials of the determinants 
\[\arraycolsep=.5ex
\left|\begin{array}{lc} X^{i,i}_{1,k} & 0 \\[1ex] X^{\varkappa,n}_{1,k}& X^{\varkappa,n}_{i,i} \end{array}\!\right|, 
\quad k\in\{1, \dots, [n/2]\},\quad i=k,\dots,\varkappa, 
\]
contain coordinate functions associated with noncommuting basis elements of the algebra $\mathfrak t_{\gamma}(n)$.
More precisely, each of the monomials includes two such coordinate functions, namely, 
$x_{ii'\!}$ and $x_{j'\!i}$ for some values $i'\in\{1,\dots,k\}$ and $j'\in\{\varkappa,\dots,n\}$. 
It is sufficient to symmetrize only the corresponding pairs of basis elements. 
As a result, after the symmetrization and the transposition of the matrices we obtain the following expression 
for the invariant of $\mathfrak t_{\gamma}(n)$ corresponding to the invariant with $x_0$ from
Theorem~\ref{TheoremOnBasisOfInvsOfCoadjRepresentationOfSolvAlgsWithTriangularNilradicalAnd1NonnilpElement}:
\[
f+\sum_{k=1}^{\left[\frac n2\right]} \frac{(-1)^{k+1}}{|\mathcal E^{1,k}_{\varkappa,n}|}
(\gamma_k-\gamma_{k+1}) \sum_{k<i<\varkappa}\sum_{i'=1}^k\sum_{j'=\varkappa}^n
\frac{e_{i'\!i}e_{ij'\!}+e_{ij'\!}e_{i'\!i}}2(-1)^{i'\!j'}\bigl|\mathcal E^{1,k;\hat i'}_{\varkappa,n;\hat j'}\bigr|.
\]
Here $\bigl|\mathcal E^{1,k;\hat i'}_{\varkappa,n;\hat j'}\bigr|$ denotes 
the minor of the matrix $\mathcal E^{1,k}_{\varkappa,n}$ complementary to the element $e_{i'\!j'\!}$. 
Since $e_{i'\!i}e_{ij'\!}=e_{ij'\!}e_{i'\!i}+e_{i'\!j'\!}$, then 
\[
\sum_{i'=1}^k\sum_{j'=\varkappa}^n
\frac{e_{i'\!i}e_{ij'\!}+e_{ij'\!}e_{i'\!i}}2(-1)^{i'\!j'}\bigl|\mathcal E^{1,k;\hat i'}_{\varkappa,n;\hat j'}\bigr|=
\arraycolsep=.5ex
\left|\begin{array}{lc} \mathcal E^{1,k}_{i,i} & \mathcal E^{1,k}_{\varkappa,n} \\[1ex]
0 & \mathcal E^{i,i}_{\varkappa,n} \end{array}\!\right|
\pm\frac12|\mathcal E^{1,k}_{\varkappa,n}|, 
\]
where we have to take the sign `$+$' (resp. `$-$') if 
the elements of~$\mathcal E^{1,k}_{i,i}$ are placed after (resp. before) 
the elements of~$\smash{\mathcal E^{i,i}_{\varkappa,n}}$ in all the relevant monomials. 
Therefore, up to a constant summand we derive 
the expression for the last element of the invariant basis given in case 1 of the statement. 
It is formally obtained from the corresponding expression in $x$'s
by the replacement $x_{ij}\to e_{ji}$ and $x_0\to f$ and the transposition of all the matrices. 
That is why we assume that the symmetrization procedure is trivial in the sense described.
Let us emphasize that 
a uniform order of elements from $\mathcal E^{1,k}_{i,i}$ and $\mathcal E^{i,i}_{\varkappa,n}$ 
has to be fixed in all the monomials under usage of the `non-symmetrized' form of invariants. 
\end{proof}

\begin{corollary}\label{CorollaryOnRationalInvsOfSolvAlgsWithTriangularNilradicalAnd1NonnilpElement}
If $\gamma_k=\gamma_\varkappa$ for all $k\in\{1,\dots,[n/2]-1\}$, then $\Inv(\mathfrak t_\gamma(n))$ has a basis from Casimir operators.
Otherwise, the algebra~$\mathfrak t_\gamma(n)$ admits a rational basis of invariants if and only if
$\alpha_k\in\mathbb Q$ for all $k\in K:=\{k_0+1,\dots,[n/2]\}$,
and it admits a polynomial basis of invariants if and only if additionally 
either $\alpha_k=0$ for all $k\in K$ or $\alpha_k>0$ for some $k\in K$.
Here $k_0$ is the minimal value of $k$ for which $\gamma_k\ne\gamma_\varkappa$.
\end{corollary}

\begin{remark}\label{RemarkOnAlgReformulationOfCondOfExtensionOfInvSetForSolvAlgsWithTriangularNilradicalAnd1NonnilpElement}
It follows from Theorem~\ref{TheoremOnBasisOfInvsOfSolvAlgsWithTriangularNilradicalAnd1NonnilpElement} that
the maximal number $N_{\mathfrak t_\gamma(n)}$ of functionally independent invariants
of the algebra $\mathfrak t_\gamma(n)$ is equal to $[n/2]+1$
if $\gamma_k=\gamma_\varkappa$ for all $k\in\{1,\dots,[n/2]\}$
and to $[n/2]-1$ otherwise.
The condition on the extension of~$\Inv(\mathfrak t_\gamma(n))$ can be reformulated in terms of commutators
in the following way:
The nilindependent basis element~$f$ commutes with the `nilpotent' basis elements
$e_{k\varkappa}$, $k=1,\dots,[n/2]$, lying on the significant part of the secondary diagonal
of the basis `matrix'~$\mathcal E$, i.e., $[f,e_{k\varkappa}]=0$, $k=1,\dots,[n/2]$.
\end{remark}

\begin{remark}
The significant elements of the secondary diagonal of the lifted invariant matrix play
a singular role under the normalization procedure in all investigated algebras with
nilradicals isomorphic to $\mathfrak t_0(n)$:
$\mathfrak t_0(n)$ itself and $\mathfrak{st}(n)$~\cite{Boyko&Patera&Popovych2007b} as well as
$\mathfrak t_\gamma(n)$, which is studied in this paper.
(More precisely, in~\cite{Boyko&Patera&Popovych2007b} the normalization procedure was realized
for $\mathfrak t(n)$ and then the results on the invariants were extended to $\mathfrak{st}(n)$.)
The reasons for such a singularity were not evident from the consideration of~\cite{Boyko&Patera&Popovych2007b}.
Only Remark~\ref{RemarkOnAlgReformulationOfCondOfExtensionOfInvSetForSolvAlgsWithTriangularNilradicalAnd1NonnilpElement} 
gives an explanation for this and justifies the naturalness of the chosen normalization conditions.
\end{remark}

\section{Conclusion and discussion}

Using the technique developed in~\cite{Boyko&Patera&Popovych2007b} for triangular algebras
in the framework of our original pure algebraic approach~\cite{Boyko&Patera&Popovych2006,Boyko&Patera&Popovych2007},
in this paper we investigated the invariants of solvable Lie algebras with nilradicals isomorphic to
$\mathfrak t_0(n)$ and one nilindependent `diagonal' element.
The algorithm has two main steps. They are constructed from explicit formulae for a fundamental
lifted invariant of the coadjoint representation of the corresponding connected Lie group
and the normalization procedure for excluding parameters from lifted invariants.
Realization of both steps for the algebras under consideration are more difficult than for
the universal triangular algebras $\mathfrak t_0(n)$ and $\mathfrak t(n)$.
Thus, a fundamental lifted invariant has a more complex representation.
One of its component does not admit a good interpretation as an element of
the matrix of the significant part of which is formed by the other components.
The choice of normalization conditions essentially depends on the algebra parameters
that lead to the furcation of the calculations and final results.

There are two principally different cases on the number of normalization conditions
and, therefore, on the cardinality of the fundamental invariants.
If $\gamma_k=\gamma_\varkappa$ for all $k\in\{1,\dots,[n/2]\}$ (the singular case),
the algebra $\mathfrak t_\gamma(n)$ has $[n/2]+1$ functionally independent invariants.
The basis of $\Inv(\mathfrak t_\gamma(n))$, constructed in
Theorem~\ref{TheoremOnBasisOfInvsOfSolvAlgsWithTriangularNilradicalAnd1NonnilpElement} for this case,
consists of polynomial invariants forming a basis of $\Inv(\mathfrak t_0(n))$
and one more nominally rational invariant which includes the chosen nilindependent element~$f$,
and can be replaced by a more complicated polynomial invariant.
Otherwise (the regular case),
the maximal number $N_{\mathfrak t_\gamma(n)}$ of functionally independent invariants
of the algebra $\mathfrak t_\gamma(n)$ is equal to $[n/2]-1$.
In this case a basis of $\Inv(\mathfrak t_\gamma(n))$ can be presented via combinations of powers of the 
basis invariants of $\Inv(\mathfrak t_0(n))$.
The basis is polynomial or rational only under special restrictions on the algebra parameters.
The conjecture of~\cite{Tremblay&Winternitz2001} on the number and form of elements in the bases
is corroborated.
Only in the regular case should the basis be written more precisely.

In spite of the above difficulties, the calculations are quite handy due to the use of the optimized technique.
This technique includes
the choice of special coordinates in the inner automorphism group,
the matrix representation of most of the lifted invariants
and the natural normalization constraints associated with the algebra structure.
The cardinality of the invariant basis is determined in the process of finding the invariants.
Moreover, we only partially constrain the lifted invariants in the beginning of the normalization procedure.
The total number of necessary constraints and any additional constraints are specified before the completion of the normalization.
As a result of the optimization, eliminating of the group parameters in the singular case is reduced to
a linear system of (algebraic) equations.
After solving a similar linear system in the regular case, we eliminate most of the group parameters and
obtain nonlinear algebraic equations for the elimination of only one parameter, these equations are trivial.

The present investigation can be directly extended to similar solvable Lie algebras with more
nilindependent diagonal elements. All such algebras are embedded in $\mathfrak{st}(n)$ as ideals.
The technique should be modified slightly.
An entirely different matter is the investigation of the other solvable Lie algebras with nilradicals isomorphic to
$\mathfrak t_0(n)$.
It is not yet known whether we will be able to use the partial matrix representation of the
lifted invariants, as well as other tricks lifted from the technique explained herein, as applied to this problem.

\section*{Acknowledgments}
The work was partially supported by the National Science and Enginee\-ring
Research Council of Canada, by the MIND Institute of Costa Mesa, Calif., and by MITACS.
The research of R.\,P. was supported by Austrian Science Fund (FWF), Lise Meitner project M923-N13 and project P25064. 
V.\,B. is grateful for the hospitality extended to him at the Centre de Recherches Math\'ematiques, Universit\'e de Montr\'eal. 
The authors thank the referees for useful remarks.

\end{document}